\documentclass[prl,twocolumn,showpacs]{revtex4}
\usepackage{graphicx}
\usepackage{graphicx}    
\usepackage{dcolumn}     
\usepackage{bm}          
\def\etal{{\it et\ al.\ }}

\newcommand{\lsim}
 {\ \raise.35ex\hbox{$<$}\kern-0.75em\lower.5ex\hbox{$\sim$}\ }
\newcommand{\gsim}
 {\ \raise.35ex\hbox{$>$}\kern-0.75em\lower.5ex\hbox{$\sim$}\ }
\def\journal #1#2#3#4{#1 {\bf #2}, #3 (#4)}
\def\PR{Phys.\ Rev.}
\def\PRB{Phys.\ Rev.\ B}
\def\PRL{Phys.\ Rev.\ Lett.}

\def\IJMP{Int.\ J.\ Mod.\ Phys.}

\def\JPSJ{J.\ Phys.\ Soc.\ Jpn.}

\def\PTP{Prog.\ Theor.\ Phys.}

\begin{document}
\preprint{Hubbard kinetic}

\title{Crossover of superconducting properties and kinetic-energy gain 
in two-dimensional Hubbard model
}
\author{
Hisatoshi Yokoyama$^{1}$, Yukio Tanaka$^{2}$, Masao Ogata$^{3}$ 
and Hiroki Tsuchiura$^{4}$
}
\affiliation{
 $^{1}$Department of Physics, Tohoku University,
 Sendai 980-8578, Japan\\
 $^{2}$Department of Applied Physics, Nagoya University, 
 Nagoya 464-8603, Japan\\
 $^{3}$Department of Physics, University of Tokyo, Bunkyo-ku, 
 Tokyo 113-0033, Japan\\
 $^{4}$International School for Advanced Studies (SISSA),
 Via Beirut 2-4 34014 Trieste, Italy
} 

\date{\today}

\date{\today}
\begin{abstract} 
Superconductivity in the Hubbard model on a square lattice 
near half filling is studied using an optimization (or correlated) 
variational Monte Carlo method. 
Second-order processes of the strong-coupling expansion are 
considered in the wave functions beyond the Gutzwiller factor. 
Superconductivity of d$_{x^2-y^2}$-wave is widely stable, and 
exhibits a crossover around $U=U_{\rm co}\sim 12t$ from a BCS 
type to a new type. 
For $U\gsim U_{\rm co}$ ($U\lsim U_{\rm co}$), the energy gain 
in the superconducting state is derived from the kinetic (potential) 
energy. 
Condensation energy is large and $\propto \exp(-t/J)$ [tiny] on the 
strong [weak] coupling side of $U_{\rm co}$. 
Cuprates belong to the strong-coupling regime. 
\par

\end{abstract}
\pacs{74.20.Mn, 71.10.Fd, 71.30.+h}
\maketitle

In a superconducting (SC) transition, conventional BCS superconductors 
follow a low-frequency sum rule of optical conductivity 
$\sigma_1(\omega)$ \cite{Tinkham}. 
However, recent experiments have shown that cuprate superconductors 
violate this sum rule \cite{Basov}.
This violation implies a gain in kinetic energy ($K$) in the 
transition, because the sum of $\sigma_1(\omega)$ is proportional to 
$-K$ \cite{Maldague}.
Such kinetic-energy-driven superconductivity \cite{Hirsch} sharply 
contrasts with that of the conventional BCS superconductors, where 
the transition is induced by the lowering of potential energy 
\cite{Chester}. 
\par

The Hubbard model on a square lattice, 
${\cal H}={\cal H}_t+{\cal H}_U
         =-t\sum_{\langle i,j\rangle\sigma} 
         \left(c^\dagger_{i\sigma}c_{j,\sigma} 
              +c^\dagger_{j\sigma}c_{i,\sigma}\right) 
         +U\sum_jn_{j\uparrow}n_{j\downarrow},$
is often used as a simple model which probably seizes the essence 
of cuprates \cite{Anderson}. 
In spite of its importance, reliable knowledge is limited 
particularly in the intermediate and strong coupling regimes. 
It is still controversial whether superconductivity is realized 
in this model. 
In the strong coupling region, the Hubbard model is mapped to 
$t$-$J$-type models, where the d$_{x^2-y^2}$-wave superconductivity
is concluded by exact diagonalization \cite{Dagotto} and variation 
methods \cite{VMCtJ,YO}.
On the other hand, in the weak-coupling region, many quantum 
Monte Carlo (QMC) studies \cite{QMC} came to negative conclusions 
for $U/t=2$-$4$.
Unfortunately, QMC is ineffective in larger-$U$ regimes due to 
the negative sign problem. 
In contrast, RPA calculations \cite{RPA}, fluctuation exchange 
approximations \cite{FLEX}, renormalization-group \cite{RG1,RG2} and 
perturbative \cite{Hlubina,Kondo} studies concluded d$_{x^2-y^2}$-wave 
superconductivity.
Besides, variational Monte Carlo (VMC) studies using the Gutzwiller 
projection argued that an antiferromagnetic (AF) order prevails 
widely near half filling, and narrows the SC region \cite{Yamaji,full}. 
\par

The purpose of this letter is to resolve the above discrepancy, 
and to explore the possibility of the kinetic-energy-driven 
superconductivity in the two-dimensional Hubbard model.
By carefully studying the wave functions with vital improvement 
on the Gutzwiller projection, it is found that the d-wave SC state
is stabilized even in the weak coupling region, but its energy gain 
is too small to be observed in QMC. 
Furthermore, we find a crossover at $U_{\rm co}\sim 12t$, over which 
the SC transition is induced by the lowering of kinetic energy. 
This indicates that the high-$T_{\rm c}$ superconductivity should 
be understood in the context of strong correlation. 
\par

A VMC method \cite{YS1} is useful to our purpose for its applicability 
to any $U/t$. 
Generally, a Jastrow-type function, $\Psi={\cal P}\Phi$, is used as 
a trial state, where $\Phi$ signifies a one-body (Hartree-Fock) state, 
and ${\cal P}$ a correlation factor.
For ${\cal P}$, the Gutzwiller (onsite) factor \cite{GW}, 
${\cal P}_{\rm G}=\prod_j[1-(1-g)n_{j\uparrow}n_{j\downarrow}]$, 
has been often chosen for its simplicity. 
Although ${\cal P}_{\rm G}$ is successful in $t$-$J$-type 
models \cite{GJR,YO}, it brings about unfavorable results for 
the Hubbard model \cite{YS1}, which were expected to be remedied by
improving ${\cal P}_{\rm G}$ \cite{improve,YS3}. 
Since the relationship between the Hubbard and $t$-$J$-type models is, 
${\cal H}_{t-J}\sim e^{iS}{\cal H}_{\rm Hub}e^{-iS}$ \cite{Harris}, 
which yields
$$
\frac{\langle\Psi_{\rm G}|{\cal H}_{t-J}|\Psi_{\rm G}\rangle}
{\langle\Psi_{\rm G}|\Psi_{\rm G}\rangle}
\sim
\frac{\langle\Psi_{\rm G}e^{iS}|{\cal H}_{\rm Hub}|e^{-iS}\Psi_{\rm G}\rangle}
{\langle\Psi_{\rm G}e^{iS}|e^{-iS}\Psi_{\rm G}\rangle}, 
$$
one can make improvements by applying the strong coupling expansion, 
$e^{-iS}$, to $\Psi_{\rm G}$ ($={\cal P}_{\rm G}\Phi$). 
\par

\begin{figure}[hob]
\vspace{-0.2cm}
\begin{center}
\includegraphics[width=8.5cm,clip]{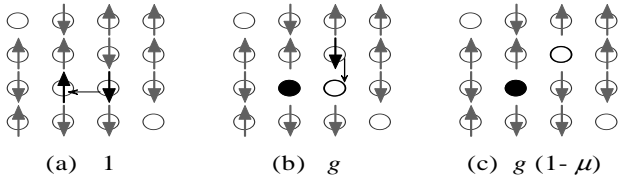}
\end{center}
\vskip -7mm
\caption{
Weight assignment of ${\cal P}_Q$. 
A closed circle indicates a doubly-occupied (d-) site.
{\bf (a)} A configuration appearing for $U/t=\infty$. 
{\bf (b)} A d-site arises by a single hopping process from (a) 
(a virtual state in the second order). 
{\bf (c)} A d-site exists, and two hopping steps are needed to reach 
from (a) (a virtual state of the fourth order). 
}
\label{fig:wf}
\end{figure}

Along this line, we employ 
$\Psi_Q=\prod_j\left[1-\mu Q_j\right]\Psi_{\rm G}={\cal P}_Q\Phi$ 
with $0\le\mu\le 1$ in this work \cite{notewf}. 
Here, $Q_j$ is an asymmetric projection 
$Q^{\rm a}_j=d_j\prod_\tau(1-e_{j+\tau})$ for less-than-half filling 
($n<1$, $n$: electron density), where $d_j=n_{j\uparrow}n_{j\downarrow}$, 
$e_j=(1-n_{j\uparrow})(1-n_{j\downarrow})$, and $\tau$ runs over all
the nearest-neighbor sites. 
This factor takes account of virtual states in the second order 
of the strong-coupling expansion, as explained in Fig.~\ref{fig:wf}.
As a one-body function $\Phi$, we use the following: 
Fermi sea $\Phi_{\rm F}$ for a normal state, 
a mean-field solution $\Phi_{\rm AF}(\Delta)$ \cite{YS2} for an 
AF state, and a fixed-$n$ BCS state with d$_{x^2-y^2}$-wave 
symmetry $\Phi_{\rm SC}(\Delta,\zeta)$ for a SC state \cite{notesymm}. 
Here, $\Delta$ and $\zeta$ are variational parameters, although they 
denote gaps and chemical potential, respectively, in the mean-field 
approximations. 
\par

Since our trial functions have up to four parameters to be optimized, 
the use of optimization VMC procedure \cite{OVMC} is practically 
indispensable. 
Special treatment for the present calculations is that after the 
convergence of optimization, we continue iteration for a while, 
and average the data obtained during this excess process. 
Thereby, the accuracy in energy is markedly increased, sometimes 
to the order of $10^{-5}t$. 
We used lattices of $L\times L$ sites ($L=6$-16) with closed shells 
and periodic-antiperiodic boundary conditions, and collected typically 
$2\times 10^5$-$10^6$ samples, keeping the acceptance ratio 0.5. 
\par
%

\begin{figure}[htb]
\begin{center}
\includegraphics[width=8.7cm,clip]{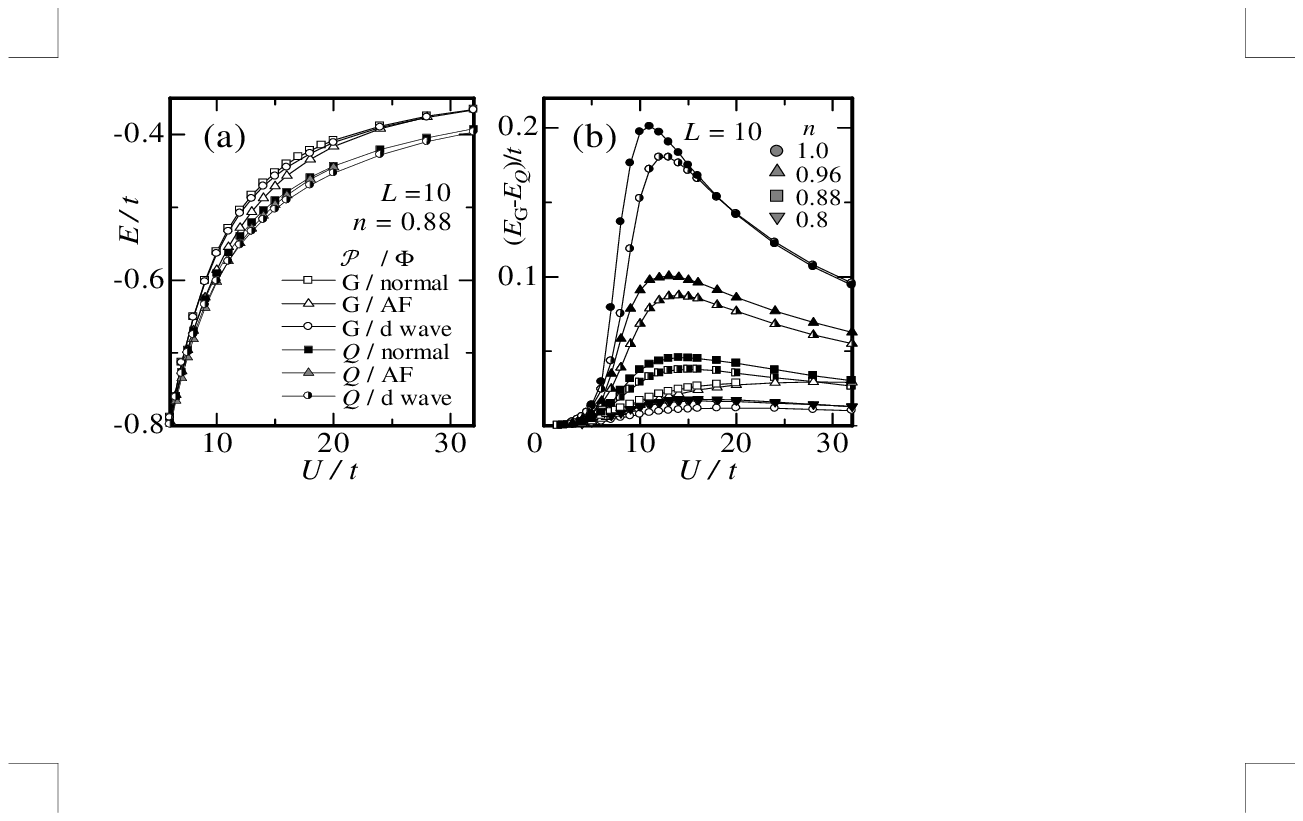}
\vskip -4mm
\caption{
{\bf (a)} Comparison of total energy among various functions.
For instance, `$Q$/normal' indicates $\Psi={\cal P}_Q\Phi_{\rm F}$. 
{\bf (b)} Energy improvement by $\Psi_Q$ on $\Psi_{\rm G}$ in each state 
for several values of $n$. 
Solid (open, half-solid) symbols indicate the d-wave (AF, normal) state. 
For $n\lsim 0.85$, the AF phase is not stabilized for any $U/t$. 
\label{fig:totE}}
\end{center}
\end{figure}

First, we compare the total energy $E/t$ between 
$\Psi_Q$ and $\Psi_{\rm G}$ [Fig.\ref{fig:totE}(a) for $n=0.88$]. 
In each state, $E/t$ is noticeably improved by the correlation 
factor ${\cal P}_Q$, particularly in large-$U$ regimes. 
In Fig.\ref{fig:totE}(b), the magnitude of the improvement is 
depicted for some values of $n$. 
The SC and normal states have large values which increase as $n$ 
approaches 1, whereas the value of the AF state is small and 
decreases as $n\rightarrow 1$. 
The improvements in the former states are caused by the lowering 
of kinetic energy $E_t=\langle{\cal H}_t\rangle$ (not shown), because 
${\cal P}_Q$ promotes the hopping between neighboring sites by 
making the configurations like Fig.\ref{fig:wf}(b) advantageous. 
In contrast, in the AF state, the binding effect between a d-site 
and an empty (e-) site is already included in $\Phi_{\rm AF}$; 
the number of broken antiparallel-spin bonds increases in a N\'eel 
background when an e-site goes away from its partner d-site. 
Thus, the effect of ${\cal P}_Q$ on $\Phi_{\rm AF}$ is small and 
reduced as $n\rightarrow 1$. 
This result shows that the correlation factor ${\cal P}_Q$ is 
essential for the normal and SC states.
Henceforth, we focus on $\Psi_Q$. 
\par

\begin{figure}[htb]
\vspace{-0.2cm}
\begin{center}
\includegraphics[width=8.5cm,clip]{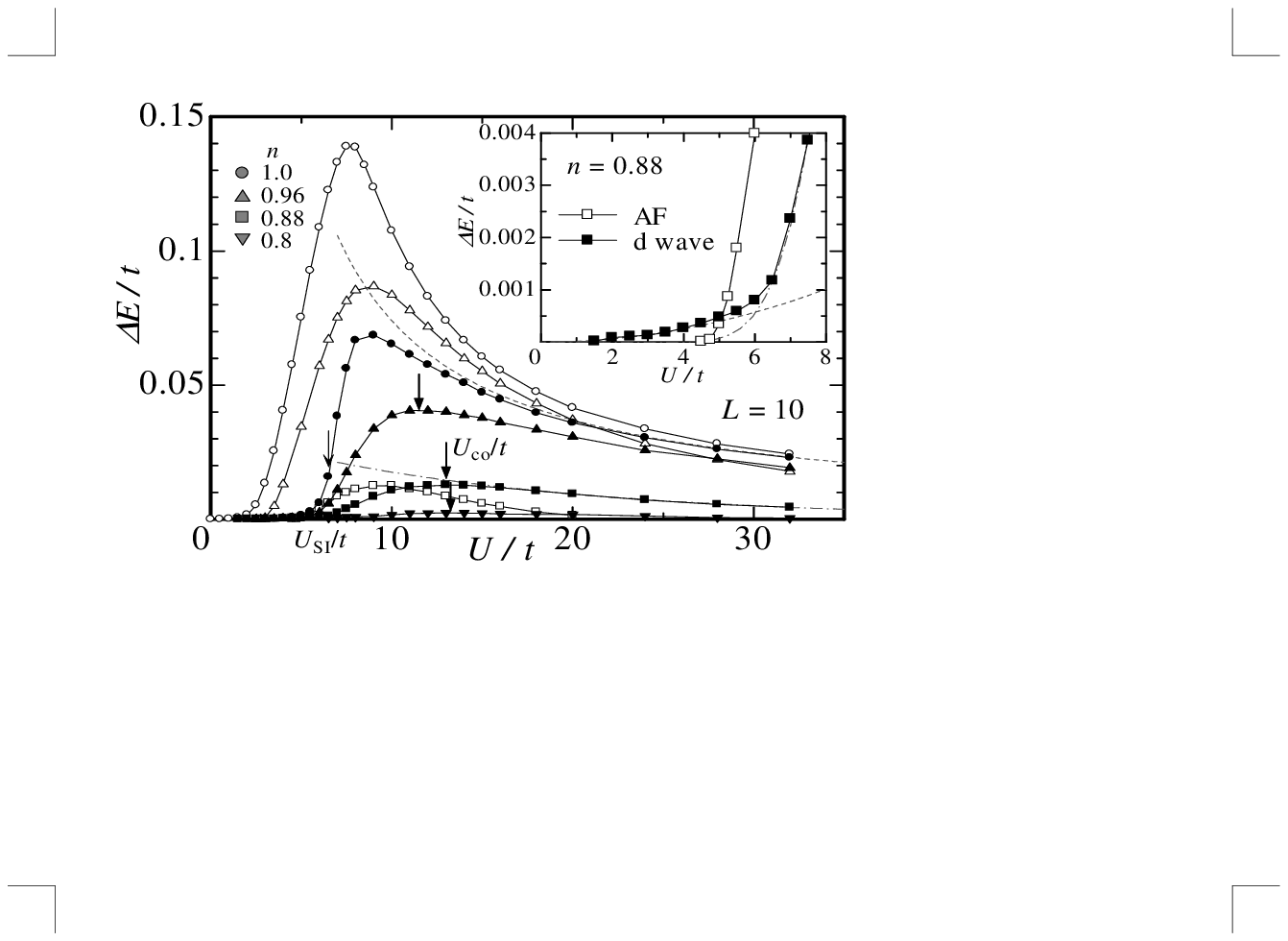}
\end{center}
\vskip -7mm
\caption{
Energy difference (condensation energy) between the d-wave (AF) 
and the normal states, estimated by $\Psi_Q$. 
Solid (open) symbols denote the SC (AF) state.
Additional lines for $n=1$ and $n=0.88$ are $\propto t/U$ (dashed), 
and $\exp(-U/t)$ (dash-dotted), respectively.
Shown in inset is magnification of the small-$U$ region for $n=0.88$. 
Additional lines are $\propto (U/t)^2$ (dashed) and 
$\exp(-t/U)$ (dash-dotted). 
An arrow for $n=1$ indicates $U_{\rm SI}/t$, and those for $n<1$ 
$U_{\rm co}/t$ at which $\Delta E$ has a maximum.
}
\label{fig:conde}
\end{figure}

\begin{figure}[hob]
\vspace{-0.4cm}
\begin{center}
\includegraphics[width=8.5cm,clip]{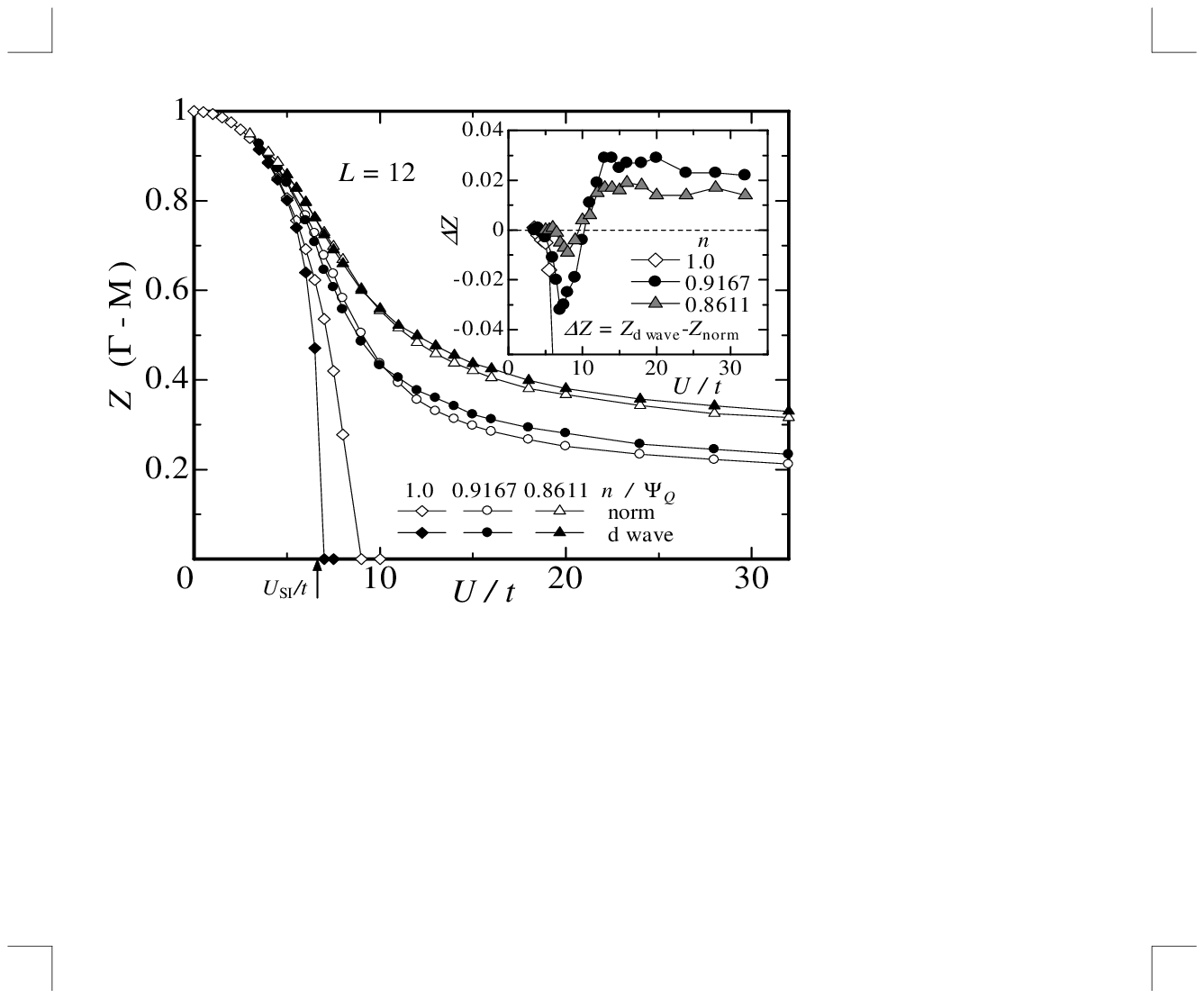}
\vskip -4mm
\caption{
Quasiparticle renormalization factor in the nodal direction for 
the normal (open) and the d-wave (solid) states. 
The inset shows the difference of $Z$ between the normal and the 
d-wave states.
\label{fig:Z}
}
\end{center}
\end{figure}

Next, we consider the energy difference between the normal and the 
ordered states (condensation energy), 
$\Delta E=E^{\rm norm}-E^{\rm order}$, which is plotted in 
Fig.\ref{fig:conde}. 
At half filling, the AF state is dominant for any value of $U$ 
as anticipated, and has a peak at $U=U_{\rm co}^{\rm AF}\sim 8t$, 
which originates in a metal-insulator (Mott) transition of the normal 
state \cite{Yatt}. 
Phase-transition behavior was found in $\Psi_Q^{\rm SC}$, 
although $\Psi_Q^{\rm SC}$ does not give the lowest variational energy 
for $n=1$. 
As shown in Fig.\ref{fig:conde}, $\Delta E$ for the SC state abruptly 
increases at $U=U_{\rm SI}\sim 6.5t$, and the increase there becomes 
sharper as $L$ increases (not shown).
Note that this behavior is inherent in the SC state, and 
$E^{\rm norm}$ is smooth near $U_{\rm SI}$. 
Around $U_{\rm SI}$, the quasiparticle renormalization factor $Z$
for $\Psi_Q^{\rm SC}$ vanishes (Fig.\ref{fig:Z}), and the small-$q$ 
behavior of the charge structure factor $N(q)$ changes from linear to 
$q^2$-like (not shown). 
Therefore, we consider that a superconductor-insulator transition 
takes place, and $\Psi_Q^{\rm SC}$ becomes insulating for $U>U_{\rm SI}$. 
Actually, the parameter $g$ exhibits a cusp, and $\Delta$ and $\mu$ 
drastically change at $U=U_{\rm SI}$. 
The insulating behavior with $\Delta E\propto t^2/U$ is consistent 
with the aspect of $t$-$J$-type models \cite{VMCtJ,YO,Randeria}. 
\par 

When $n$ deviates from half filling, the behavior of $\Delta E$ 
is no longer a transition, but becomes a crossover. 
The inset of Fig.\ref{fig:conde} shows a close-up of $\Delta E$ 
of the small-$U$ region for $n=0.88$. 
It is found that $\Delta E$ is very small for $U\lsim 6.5t$ 
($\sim 10^{-4}t$ at $U=4t$), and behaves mildly like a power-law 
function \cite{notesmallU}. 
It is probable that due to this imperceptibly weak superconductivity, 
many QMC studies for $U/t=2$-4 \cite{QMC} overlooked signs of 
SC order. 
At $U\sim 6.5t$, an exponential-like rapid increase occurs which 
again originates in $\Psi_Q^{\rm SC}$, and then $\Delta E$ reaches 
a maximum at $U=U_{\rm co}$ as indicated by arrows in Fig.\ref{fig:conde} 
\cite{noteco}. 
As we will explain shortly, the SC properties for 
$U>U_{\rm co}$ qualitatively differ from those for $U<U_{\rm co}$. 
For $U>U_{\rm co}$, $\Delta E$ is nicely fitted by the form $\exp(-U/t)$ 
[or $\exp(-t/J)$], which means that the effective attractive interaction 
is $J=4t^2/U$; a viewpoint from $t$-$J$-type models is justified. 
In this region, the SC state is fairly stable to considerably 
large $U/t$. 
As $n$ decreases, the difference between the two regimes tends to be 
vague. 
\par
\begin{figure}[htb]
\begin{center}
\includegraphics[width=8.5cm,clip]{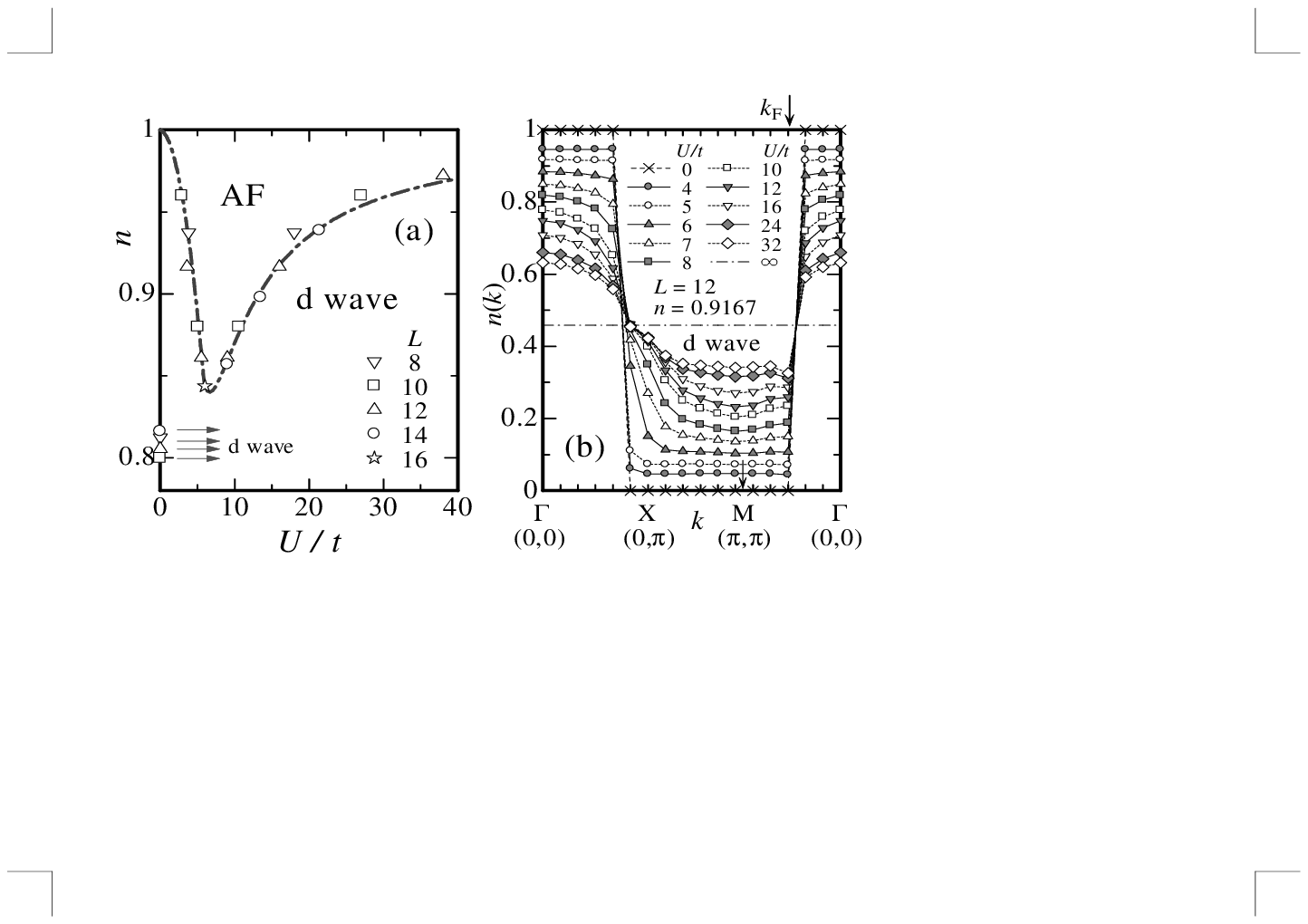}
\end{center}
\vskip -6mm
\caption{
{\bf (a)} Phase diagram within $\Psi_Q$ near half filling. 
The boundary for $U/t\lsim 6$ is well fitted by $1-n\propto (U/t)^2$, 
which looks similar to the result of a renormalization-group 
theory \cite{RG1}. 
For systems with $n<0.84$, the d wave is always stable for $U/t>0$, 
as illustrated by gray arrows.
{\bf (b)} Momentum distribution $n(k)$ of the d-wave state. 
M point is indicated by an arrow.
}
\label{fig:phased}
\end{figure}

Before the SC properties, let us discuss the phase diagram. 
As shown in the inset of Fig.\ref{fig:conde}, due to the sudden increase 
of $\Delta E^{\rm AF}$ for $n<1$, the stable phase switches from SC to AF, 
but the SC phase is retrieved at a larger value of $U$ (Fig.\ref{fig:conde}). 
The obtained phase diagram in the $U/t$-$n$ plane is shown in 
Fig.\ref{fig:phased}(a). 
The AF state is stabilized near half filling where the nesting 
condition is satisfied, and occupies the maximum range of $n$ at 
$U\sim 6t$. 
If we use $\Psi_{\rm G}$ instead of $\Psi_Q$, the AF region 
becomes larger. 
This implies that the improvement by $e^{-iS}$ is more effective 
on the SC state than on the AF state. 
We expect that the higher-order improvements enlarge the SC region. 
\par

\begin{figure}[hob]
\begin{center}
\includegraphics[width=8.5cm,clip]{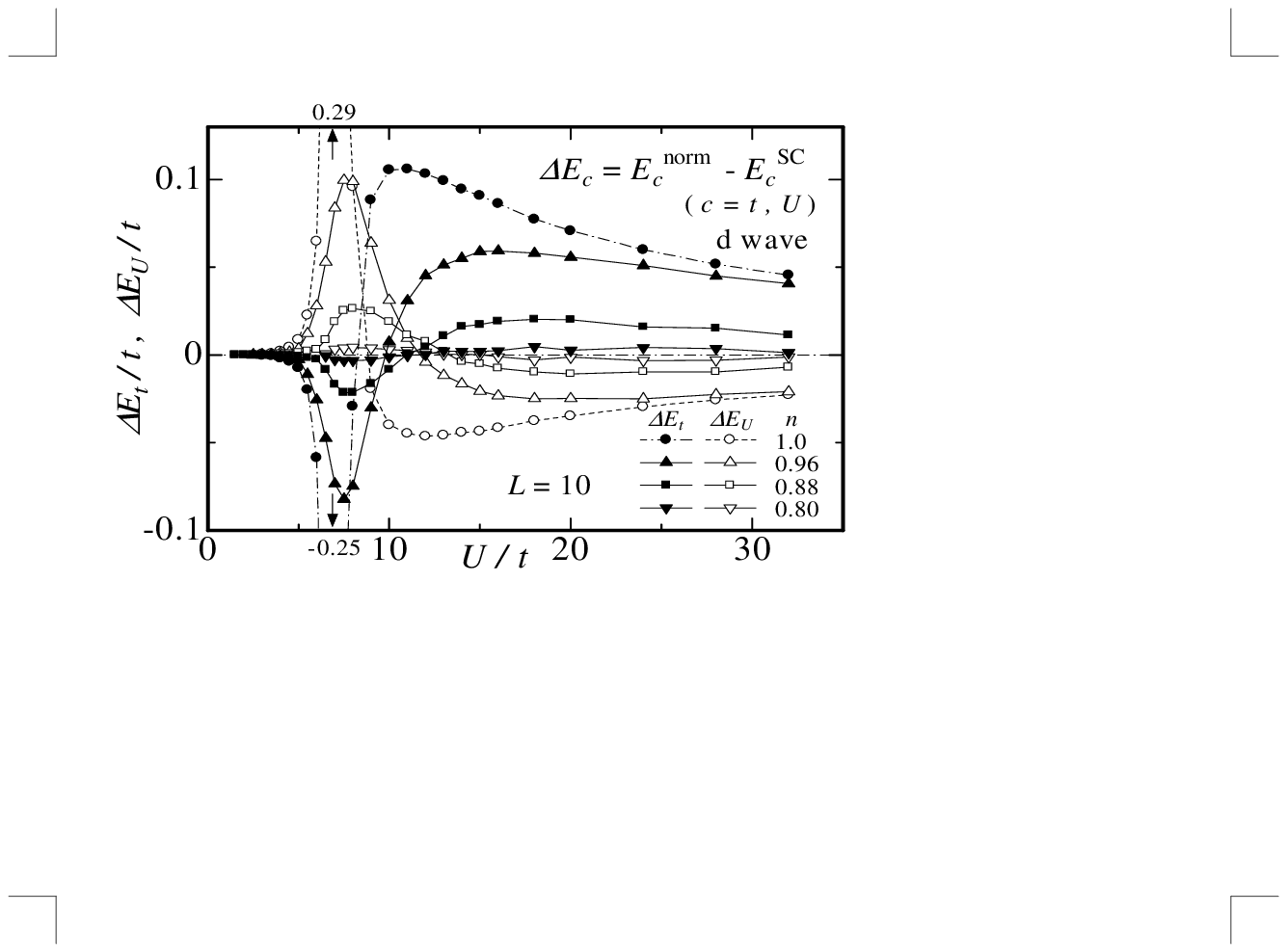}
\vskip -4mm
\caption{
Difference of kinetic and potential energies between 
the normal and SC states for several values of $n$. 
The absolute values tend to increase, as $L$ increases (not shown). 
\label{fig:kinetic}
}
\end{center}
\end{figure}

We turn to the SC properties in $\Psi_Q^{\rm SC}$. 
Figure \ref{fig:kinetic} shows the difference of kinetic energy $\Delta E_t$ 
($=E^{\rm norm}_t-E^{\rm SC}_t$) and potential energy $\Delta E_U$. 
Apparently, the mechanism of energy gain changes at $U\sim U_{\rm co}$.
For $U\lsim U_{\rm co}$, the energy gain is due to the lowering of 
$E_U$ ($\Delta E_U>0$), as in the BCS superconductors. 
Inversely, for $U\gsim U_{\rm co}$, kinetic-energy gain occurs 
($E_t$ is lower in the SC state), agreeing with cuprates. 
This qualitative change is a crossover and not a phase transition.
These aspects do not depend on $n$ qualitatively, although the magnitude 
of both $\Delta E_t$ and $\Delta E_U$ decreases, as $n$ decreases. 
The lowering of $E_t$ has been found also by a dynamical 
cluster approximation \cite{DCA}. 
\par


The momentum distribution function $n(k)$ gives another evidence for
the crossover. 
Since there are nodes in the d$_{x^2-y^2}$-wave state, $n(k)$ has a 
discontinuity at $k=k_{\rm F}$  in the $y=\pm x$ ($\Gamma$-M) direction 
[Fig.\ref{fig:phased}(b)]. 
The magnitude of this jump represents the renormalization factor $Z$, 
which is related to the inverse of the electron effective mass. 
The estimated $Z$ for $\Psi^{\rm norm}$ and $\Psi^{\rm SC}$ 
are plotted in Fig.\ref{fig:Z}. 
For $U\lsim U_{\rm co}$, the normal state has a larger value than 
the SC state, whereas this relation is reversed for 
$U\gsim U_{\rm co}$ (see also inset). 
For $U>U_{\rm co}$, electrons are hard to move (or heavy) in the 
normal state, hampered by the strong correlation. 
On the other hand, in the SC state, the coherence somewhat relieves 
the suppressed mobility, and thus the energy gain occurs in $E_t$. 
\par

This kind of crossover from weak- to strong-coupling regimes is not 
restricted to the present case. 
The attractive Hubbard model exhibits similar behavior in the s-wave 
SC transition for any electron density \cite{Yatt}. 
Hence, the kinetic-energy-driven superconductivity arises, not because 
the model has repulsive interaction, but the interaction is strong enough. 
The VMC calculations for the $t$-$J$ model showed that the exchange 
(hopping) energy is lowered (raised) in the d$_{x^2-y^2}$-wave SC state 
(See Table I in ref.\cite{YO}, and also \cite{t-J}).
Since the kinetic part of the Hubbard model induces the exchange term 
($J=4t^2/U$), the results of $t$-$J$-type models are compatible with 
the present results in the Hubbard model for $U\gsim U_{\rm co}$. 
\par

The magnitude of condensation energy in our calculation is larger than 
that obtained experimentally \cite{CE}. 
We anticipate that further refinement on the wave functions, especially 
on $\Psi^{\rm norm}$, leads to a better outcome. 
\par

In summary, we have studied the two-dimensional Hubbard model near 
half filling, based on evolved VMC calculations. 
It is shown that the d$_{x^2-y^2}$-wave SC state is stabilized in 
the Hubbard model if the variational states are improved. 
Furthermore, we find a crossover at around $U_{\rm co}/t=11$-$13$. 
For $U\lsim U_{\rm co}$, the SC state is similar to the conventional 
BCS state, but for $U\gsim U_{\rm co}$, the SC transition is caused 
by lowering the kinetic energy. 
Recent experiments of $\sigma_1(\omega)$ for cuprates \cite{Basov} 
accord with the characteristics of this strong-coupling region. 
Extended description will be given elsewhere \cite{full}.  
\par

The authors thank D.S.~Hirashima for helpful comment.
This work is partly supported by Grant-in-Aids from 
the Ministry of Education etc.~of Japan, and by 
the Supercomputer Center, ISSP, University of Tokyo.


\begin{thebibliography}{99}
\bibitem{notewf}
For half filling, a symmetric projection between $d_j$ and $e_j$, 
$Q^{\rm s}_j=d_j\prod_\tau(1-e_{j+\tau})+e_j\prod_\tau(1-d_{j+\tau})$, 
was introduced in ref.\cite{YS3}. 
We use $Q^{\rm s}_j$ also in this work for $n=1$, because 
$Q^{\rm s}_j$ is basically the same with $Q^{\rm a}_j$ for 
$n=1$ and gives a better energy for large $U/t$ than $Q^{\rm a}_j$. 
For $n\ne 1$ and $U>0$, $Q^{\rm s}_j$ does not work.
\bibitem{notesymm} We calculated energies also for homogeneous 
and extended s waves. 
Since they do not reduce energy near half filling, here we 
concentrate on a d$_{x^2-y^2}$ wave.
\bibitem{notesmallU} 
For $U\rightarrow 0$, a perturbation study of the BCS 
function\cite{Kondo} reached the result, $\Delta\propto\exp(-t^2/U^2)$. 
Such a singular term should exist, but probably is dominant 
only for extremely small $U$. 
In fact, the energy reduction by d$_{x^2-y^2}$-wave without a 
projection factor (literally BCS function) is, if any, of order 
$10^{-5}t$ for $U/t=1$ ($0.8\le n\le 1.0$).
\bibitem{noteco} 
The value of $U_{\rm co}/t$ is 
about 11 ($n=0.97$), 12 ($0.92$) and 13 (0.86 and 0.81) for $L=12$. 
Further improvement of $\Psi$ will shift $U_{\rm co}$ to a somewhat 
smaller value.
\end{thebibliography}
\end{document}